\documentclass[lettersize,journal]{IEEEtran}
\usepackage{amsmath,amsfonts}
\usepackage{mdframed}
\usepackage{xcolor}
\newmdenv[linewidth=1pt, innerleftmargin=10pt, innerrightmargin=10pt, innertopmargin=10pt, innerbottommargin=10pt]{defbox}
\newtheorem{definition}{Definition}
\usepackage{algorithmic}
\usepackage{algorithm}
\usepackage{subcaption}
\usepackage[justification=centering]{caption}
\usepackage{listings}
\usepackage{tikz}
\usepackage{textcomp}
\usepackage{stfloats}
\usepackage{url}
\usepackage{verbatim}
\usepackage{graphicx}
\usepackage{cite}
\usepackage{bm}
\begin{document}

\title{Towards Reinforcement Learning for Exploration of Speculative Execution Vulnerabilities}

\author{Evan Lai$^\dag$, Wenjie Xiong$^\ddag$, Edward Suh$^\$$, Mohit Tiwari$^\dag$, Mulong Luo$^\dag$\\

$\dag$ UT Austin, $\ddag$ Virginia Tech, $\$$ Cornell\\
\{laievan,mulong\}@utexas.edu, wenjiex@vt.edu, suh@ece.cornell.edu, tiwari@austin.utexas.edu
}


\markboth{}%
{Shell \MakeLowercase{\textit{et al.}}: A Sample Article Using IEEEtran.cls for IEEE Journals}


\maketitle
\begin{abstract}
Speculative execution attacks such as Spectre can be used to bypass the security isolation and steal information from other programs. Exploring speculative execution attacks on existing processors requires intensive manual reverse engineering and intimate knowledge of the processor. This reverse engineering-based approach requires extensive human effort,  which is slow and not scalable.
In this paper, we introduce SpecRL, a framework that utilizes reinforcement learning to explore speculative execution leaks in commercial-of-the shelf microprocessors. This reinforcement learning agent approach requires less reverse engineering effort while still be able to identify speculative execution vulnerabilties. Compared to existing fuzzing-based method \cite{oleksenko2022revizor}, it takes less time when the program size is large.
\end{abstract}


\section{Introduction and Background}
Speculative execution is a powerful technique implemented in the microarchitecture of many modern out-of-order processors to enhance performance. By speculatively execute instructions instead of stalling, more instructions can be executed in the processor pipeline, which improves instruction-level parallelism. One example of the performance benefit of speculative execution can be seen in conditional branches.
In conditional branches, the instruction following a conditional branch is only known after the conditional branch has been resolved. In the case of an in-order pipelined processor, the processor must stall, which is a big hit to performance. Instead, an out-of-order processor \textit{speculatively} executes the instructions following the conditional branch by \textit{predicting} the outcome of the branch, the processor can get work done while waiting for the branch to be resolved. If the wrong path was predicted, the processor flushes the pipeline and starts executing the correct path. Instructions that are executed during a mispredicted path are called \textit{transient instructions}, and the execution of transient instructions is called \textit{transient execution}.

While speculative execution and branch prediction are good for processor performance, they also create a critical security vulnerability. Transient executed instructions are problematic, as they can make unintended accesses to secrets in protected memory. From an instruction set architecture (ISA) level of abstraction, this is legit, as transient execution leaves no trace in the architectural state. From a microarchitectural view of the processor, however, transient execution leaves traces in various parts of the microarchitectural state, and can thus be transmitted through a variety of microarchictectural side-channels. This vulnerability has been exploited most notably with the Spectre \cite{Kocher2018spectre} attack in 2018.

\noindent\textbf{Identifying speculative vulnerabilities:} identification of vulnerability for speculative execution leaks can be done both pre-silicon and post-silicon. 

For pre-silicon identification \cite{hur2022specdoctor,ghaniyoun2021introspectre,fadiheh2020formal}, vulnerability identification is relatively straightforward, as the detector deals with the processor's RTL description, and thus has complete knowledge and control of the entire microarchitectural state. 

For post-silicon identification \cite{oleksenko2022revizor}, on the other hand, vulnerability identication is far more convoluted, as there is no direct way to tell whether a speculative leak has occurred. Guarnieri et. al \cite{guarnieri2020hardwaresoftwarecontractssecurespeculation} developed the hardware-software contract as a method to detect speculative leaks in black box processors through relational testing. A short definition of a speculative leak using contracts is given below.
\begin{defbox}
\begin{definition}
A speculative leak occurs when, given input $\alpha$ and input $\beta$, $$CTrace_{\alpha} = CTrace_{\beta} \text{ and } HTrace_{\alpha} \neq HTrace_{\beta},$$ where $CTrace_{i}$ and $HTrace_{i}$ are the contract and hardware traces generated given a program $P$, an input $i$, and a speculation contract $S$, and a $CTrace$ consists of a list of observations observable given a speculation contract $S$, and a $HTrace$ consists of a list of observations observable through side-channels.
\end{definition}
\end{defbox}

\noindent\textbf{Existing methods:} Spectre \cite{Kocher2018spectre} were found manually after extensive reverse engineering of the processor's behaviors. This method of identifying vulnerabilities is a slow and expensive process, as it requires a deep and expansive knowledge of both the instruction set architecture as well as the microarchitecture of a given processor. Since then, automated tools have been developed to search for speculative leaks, both at  pre-silicon and post-silicon stages. In this paper, we focus on searching for leaks  at the post-silicon stage. Existing post-silicon method such as Revizor \cite{oleksenko2022revizor}, or Medusa \cite{moghimi2020medusa}, rely on a fuzzing-based approach in order to generate test cases. These methods, however, are limited by the huge search space and sequential nature of many speculative leaks. This is because fuzzers generate in one-shot, and scale exponentially with program size.

\noindent{\bf This paper:}  We introduce SpecRL, a novel way to explore speculative execution vulnerabilities using reinforcement learning. 
Reinforcement learning has recently been used in computer science and security problems. Unlike traditional empirical methods, which rely on expert knowledge, learning-based methods generally only rely on data and training. If there is enough data and enough computational power, it can achieve superhuman performance based on empirical knowledge, as demonstrated by Google DeepMind in 2017 \cite{silver2016mastering}. For speculative execution vulnerabilities that we can reproduce in a processor, using reinforcement learning as opposed to supervised learning saves the need for a large dataset because reinforcement learning generates the training data by running the processor directly. Additionally, reinforcement learning is inherently sequential and can also potentially perform better compared to fuzzers when searching for more complex leaks. Due to these reasons, we believe reinforcement learning has a compelling case as a potential solution for exploring speculative execution vulnerabilities.

\section{Reinforcement Learning Formulation}
Reinforcement learning typically involves an \textit{agent} and an \textit{environment}. The agent interacts with the environment through \textit{actions}, and receives feedback through \textit{observations} and \textit{rewards}. Training follows the general flow below during one "step": 
\begin{enumerate}
    \item An action is chosen by the agent from the action space. 
    \item The learning environment responds to the action, returning a corresponding observation and reward. 
    \item The agent uses the returned observation and reward to update its policy during training.
\end{enumerate}

For our application, each action represents an assembly instruction, and the agent aims to build an assembly program that will trigger a speculative contract violation. SpecRL's environment follows the Gymnasium standard \cite{towers2024gymnasium} and is formulated as such: \newline
\textbf{Action Space:} The action space is a vector of instructions $\{i_0, i_1, ... , i_{n-1}\},$ of length $n$, where $n$ is the number of instructions the agent can choose from. \newline
\textbf{Observation Space:} Due to the black box nature of our environment,  the microarchitectural state of the processor is not directly visible in the observation space.
Instead, we rely on performance counters and side-channels to help us glean helpful observations about our microarchitectural state of the processor. More specifically, we can observe, for a given instruction sequence: 
\begin{enumerate}
    \item \textit{HTrace}, implemented with Prime+Probe.
\item \textit{CTrace}, implemented with a QEMU-based simulator. 
\item \textit{Number of branch misses (\#ofBRMisses)}, implemented with the INT\_MISC.RECOVERY\_CYCLES performance counter. 
\item \textit{Number of transient micro-operations (\#ofTranUOps)}, implemented as the difference between the performance counters UOPS.ISSUED\_ANY and UOPS.RETIRED\_SLOTS.
\end{enumerate}

In order to find the effect each additional instruction has, we must iteratively observe the desired instruction sequence. More formally, given an instruction sequence $$P = \{a_0, a_1, ... , a_{n-1}\}$$ of length $n$, we produce a sequence of instruction subsequences of length $1$ to $n$, $$S = \{p_0, p_1, ... , p_{n-1}\},$$ where $p_i = \{a_0, a_1, ... , a_{i}\}$. For each instruction subsequence $p_i$, we return the associated HTrace, Ctrace, \#ofBRMisses, and \#ofTranUOps for each input, as well as the index of $a_i$ in the action space. Specifically, an instruction subsequence $$p_i \text{ gives } obs_i = (H, C, B, T),$$ where $H = \{h_1, h_2, ... ,h_{n}\},$ $C = \{c_1, c_2, ... , c_{n}\},$ $B = \{b_1, b_2, ... , b_{n}\},$ and $T = \{t_1, t_2, ... , t_{n}\},$ where $h_i$, $c_i$, $b_i$, and $t_i$ are the HTrace, CTrace, \#ofBRMisses, and \#ofTranUOps for the $i$th input, respectively. Thus, we can define the observation space as $$OBS = \{obs_0, obs_1, ... , obs_{m-1}\},$$ where $m$ is the max instruction sequence length. \newline
\textbf{Reward Function:} The reward function for SpecRL is formulated  such that the agent be rewarded for actions that trigger observable transient executions. A very large reward is given if a speculative leak occurs. A smaller, but still substantial reward is given if any misspeculation occured, and a slightly larger reward compared to the misspeculation reward is given if this misspeculation is observable through a side-channel. Corresponding negative rewards are given if the program either does not induce misspeculation, or the misspeculation is not observable. A small negative reward is also given at each step to encourage conciseness.

\noindent\textbf{Environment Details:} There are three implementation details that arise from this particular reinforcement learning formulation. 

The first detail is the infinite loops. As we are allowing the agent to sequentially add instructions that by definition should include control flow instructions, we need to check if a given step creates an infinite loop before gathering the corresponding traces. This, however, cannot be decide statically, as noted by Alan Turing with his \textit{Halting Problem} \cite{turing1936computable}. We circumvent this problem dynamically by giving the program with the added step (instruction) first to a child process, measuring how long it takes to simulate the program, and throwing away the instruction if the child process takes too long. This method also ensures that each step has a maximum trace generation time, guaranteeing that training is not stalled by an overly long loop. 

The second detail is the microarchitectural state of the processor. 
$obs_{i}$ reflects the effect of the initial microachitectural state as well as the effect of an instruction sequence. To guarantee that $obs_{i}$ is the same for the same instruction sequence, we need to make sure the microarcitectural state of the processor is the same.
The microarchitecture states include the branch predictor and the cache. By executing the instruction WBINVD and running a program filled with $50$ million conditional branches before each $obs_{i}$, we effectively clobber the branch predictors pattern history table (PHT) to a deterministic state and flush the cache, resetting the relevant microarchitectural components between different observations of the same program. 

The third detail relates to memory accesses. The instruction sequence is observed on hardware in the kernel space,  meaning all memory accesses the agent makes must be sandboxed to prevent kernel crash. 

\begin{figure*}[ht!]
   \centering
   \includegraphics[width=\linewidth, height=0.3\textheight, keepaspectratio]{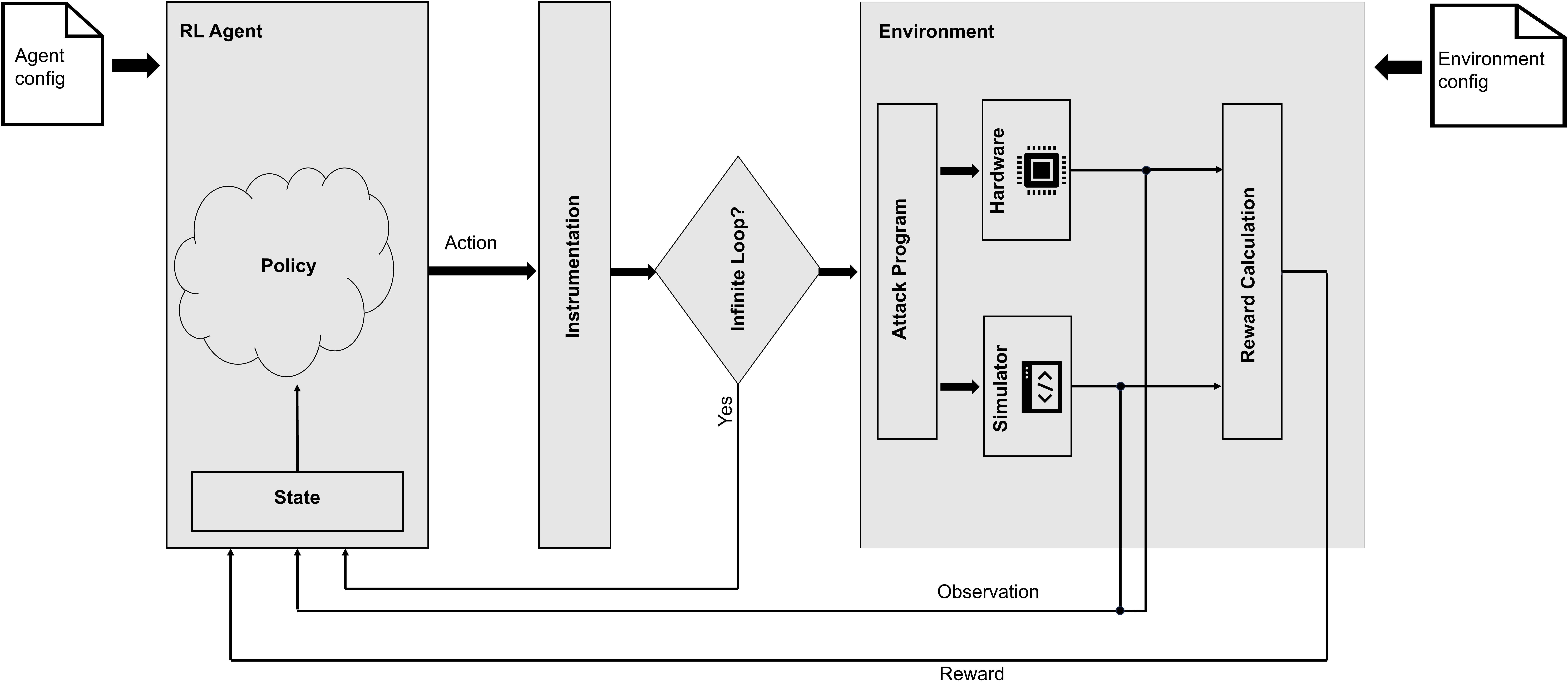}
   \caption{SpecRL's training flow}
   \label{fig:flow}
\end{figure*}

\noindent\textbf{Training Flow:} SpecRL's training flow is illustrated by Figure~\ref{fig:flow}. The agent uses its current state and policy to generate an action (instruction). This instruction is first instrumented (sandboxed) if needed. Next, an infinite loop check is done on the new instruction sequence. If the instruction sequence passes the infinite loop check, the new instruction gets added to the attack program in the environment. The attack program is then observed under a sequence of inputs to the program, either through actual hardware (for HTrace, \#ofBRMisses, and \#ofTranUOps) or through a simulator (for CTrace). These observations are fed back into the agent, and are also input into a reward function to return a reward to the agent. The agent uses these new inputs to update its policy, which then generates another action, restarting the training flow.

\section{Case Study}
The focus of this case study was to: (1) see if SpecRL could find a leak at all and (2) examine how detection time scales with program size when using SpecRL versus a existing fuzzer.
\newline
\textbf{Setup: }We implement the reinforcement learning formulation that we have specified on an AMD Matisse CPU, the AMD Ryzen 5 3600. To do this, we created a custom environment based on the Gymnasium standard, called SpecEnv, which uses Revizor \cite{oleksenko2022revizor} for testing contract violation on hardware. SpecEnv was then interfaced with Ray's RLlib, a reinforcement learning library for driving the training.

As a proof of concept,  we focused on exploiting Spectre V1 type vulnerabilities, which rely on mistraining a branch predictor. Our action space includes a simple instruction set that consists of a subtraction instruction (SBB), a signed multiplication instruction (IMUL), a conditional branch (JNS), and an unconditional branch (JMP). As there are multiple choices for each operand for each instruction (registers, labels), the action space of the agent consists of 40 different instructions. Additionally, 20 random inputs for the testing program were randomly generated at the beginning of training. Memory operation instructions are sandboxed, with all reads and writes done relative to R14, with any action that takes longer than 1 second to simulate being thrown away. 
We use RLlib's Proximal Policy Optimization (PPO) algorithm formulation to train the agent using the default algorithm configuration. 

\begin{figure}[h!]
   \centering
   \includegraphics[width=\linewidth, height=0.25\textheight, keepaspectratio]{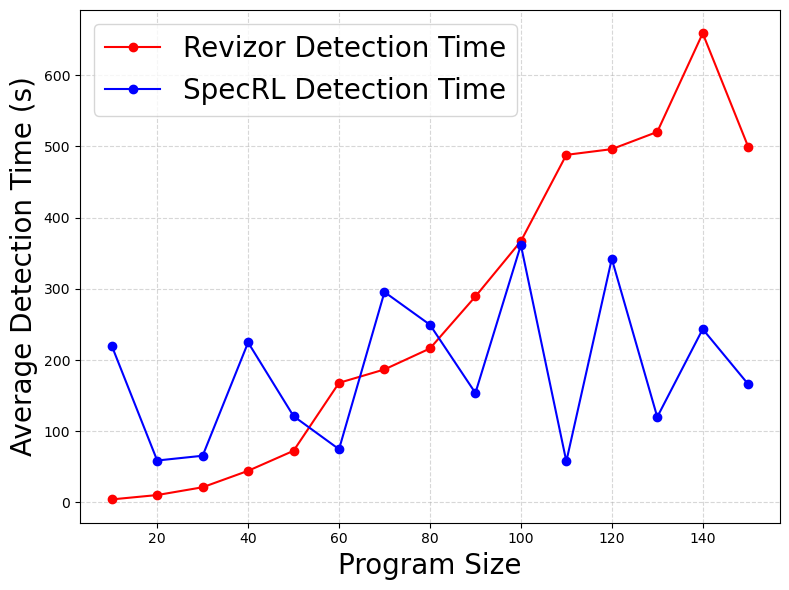}
   \caption{Average detection time as a function of program size for SpecRL and Revizor, using a simple instruction set.}
   \label{fig:DN}
\end{figure}

As a baseline for a state of the art fuzzer, we tested Revizor with a similar action space. For both SpecRL and Revizor, program size ranged from 10 to 150 in increments of 10.
\newline
\textbf{Results: } On all program sizes, SpecRL found a leak on average within 7 minutes. At larger program sizes, SpecRL was also still able to find leaks with longer instruction sequence lengths of up to 60. Additionally, as seen in Figure \ref{fig:DN}, the detection time appears only loosely correlated to program size, and at worst scales linearly, as expected. Revizor, on the other hand, has its detection time scaling roughly exponentially with program size, as seen in Figure \ref{fig:DN}. This makes sense, as given a perfectly random fuzzer, the number of test cases before a leak (which is proportional to detection time) is $\frac{a^{n-1}}{n-l+1},$ where $n$ is the program size, $a$ is the action space size, and $l$ is the length of the leak instruction sequence. These results show that SpecRL is able to find leaks more efficiently at larger program sizes.





\section{Future Work}
One of the more straightforward spaces for future work is in the action space. The case study was done with a very small action space that only consisted of 4 unique instructions. In reality, however, the x86 instruction set has hundreds of instructions. Reinforcement learning algorithms have historically struggled with large action spaces, so there are a number of challenges related to expanding the action space to a larger instruction set.

Additionally, the action space currently does not give the agent any control over the inputs that are given to the program. Currently, inputs are randomly generated, and then boosted at run time to check for speculative leaks. Although this method does work, the boosting process is quite expensive, as the program must first be analyzed to find dependencies. Including inputs in the action space could possibly speed up training, and it would also expose more of the execution environment to the agent, potentially allowing it to fully learn how to "mistrain" a branch predictor. One of the potential downsides to giving the agent control over the inputs is once again the resultant enormous action space. 

Future work could also explore enabling distributed training, which is one of the core advantages of Ray. While the current setup performs training on a single machine, leveraging the distributed nature of Ray could parallelize training and significantly accelerate the learning process. This would allow the agent to find leaks faster, and may also enable it to find more sophisticated leaks in a reasonable time.
\section{Related Work}
 Revizor \cite{oleksenko2022revizor} uses Software-Hardware Contracts, the main difference being that Revizor uses a fuzzer with some speculative filters, whereas SpecRL uses an agent, which has the concept of sequential decision making. We believe this is a more intuitive approach to speculative execution leaks that require the "training" of a branch predictor. As we have shown, Revizor does not scale well with large program sizes, while SpecRL does.


AutoCAT \cite{luo2023autocat} was the first work to explore microarchitectural attacks using reinforcement learning. It only focused on cache timing attacks, however, and does not consider the branch predictor or any speculative aspects of the microarchitecture. MACTA \cite{cui2023macta} uses multiagent RL for attack and detection of cache side-channel, but it also only considers cache timing channel mainly on a simulators.

\section{Conclusion}
In conclusion, in this paper we have introduced SpecRL, a novel method to find speculative execution leaks using reinforcement learning. We present SpecRL's reinforcement learning formulation, and we also conduct a case study examining its scalability. As speculative execution vulnerabilities continue to evolve, SpecRL gives credence to a promising direction for automating and improving microarchitectural security analysis.


\bibliographystyle{IEEEtran}
\bibliography{refs}

\end{document}